# COMBINED THERMAL CONTROL AND GNC: AN ENABLING TECHNOLOGY FOR CUBESAT SURFACE PROBES AND SMALL ROBOTS

## Salil Rabade and Jekanthan Thangavelautham

Advance in GNC, particularly from miniaturized control electronics, reaction-wheels and attitude determination sensors make it possible to design surface probes and small robots to perform surface exploration and science on low-gravity environments. These robots would use their reaction wheels to roll, hop and tumble over rugged surfaces. These robots could provide "Google Streetview" quality images of off-world surfaces and perform some unique science using penetrometers. These systems can be powered by high-efficiency PEM fuel cells that operate at 60-65 % and utilize hydrogen and oxygen electrolyzed from water. However, one of the major challenges that prevent these probes and robots from performing long duration surface exploration and science is thermal design and control. Typically there are extremes in temperature between daylight and eclipse. In the inner solar system, during the day time, there is often enough solar-insolation to keep these robots warm and power these devices, but during eclipse the temperatures falls well below the storage temperature of most devices including GNC. We have developed a thermal control system that utilizes chemicals to store and dispense heat when needed. The system takes waste products, such as water from these robots and transfers them to a thermochemical storage system. These thermochemical storage systems when mixed with water (a waste product from a PEM fuel cell power supply) releases heat. Under eclipse, the heat from the thermochemical storage system is released to keep the probe warm enough to survive. In sunlight, concentrated solar photovoltaics are used to electrolyze the water and reheat the thermochemical storage system to release the water. Our research has showed thermochemical storage systems are a feasible solution for use on surface probes and robots for applications on the Moon, Mars and asteroids.

## INTRODUCTION

Miniaturization in electronics has led to a steady reduction in form factor of spacecrafts. Thermal control has been of critical interest to the space community ever since the start of the space age in 1957. Guidance and navigation control systems including reaction wheels have operating temperatures of -40 $^o$C to 50 $^o$C. In deep space, owing to lack of solar flux, there is a need to heat these systems. This ensures the system temperature is always above the lower temperature limit .The moon and the asteroid Eros, both have lowest reported temperatures of -150 $^o$C. For robots or small sensor modules to work in this extreme environment for a sustained period, a high density energy source is needed. The high energy needs of a sensor module can be fulfilled using high efficiency PEM fuel-cells that operate at 65 % efficiency and provide an energy density up to 2,000 Wh/kg[1,13,14,15]. Fuel cells provide water as output. Typical lithium ion batteries have storage density of 120-140 Wh/kg.

A robust thermal control mechanism is needed to ensure the successful completion of any science mission. Space systems operating in the inner solar-system typically rely on solar energy and waste heat from electronics to survive the freezing temperatures. Thermal control systems can be passive or active depending on the energy source. Conventional passive control focuses on varying energy transfer rate from a system. A typical passive system does not generate energy, although it does have the capacity to do so. Active control requires solar energy to provide heat [2]. Hence, a passive thermal heating system is preferable in deep space for its simplicity and potential. The challenge is to make a passive thermal control system that can generate heat and is simple to deploy on a space system.

Thermochemical Energy Storage Systems (TESS) is extensively used on Earth to store solar energy. In this paper, a novel passive heating system is proposed using the TESS principle for space applications. This new system has the potential to be rechargeable and will only require waste cold water from fuel cells to produce heat. A low power sensor module has been developed and experiments carried out to better understand the thermal performance of a thermochemical energy storage system. The performance of the system is then compared to conventional electrical heaters. First, we review related work on sensor networks for space and the associated thermal control problems, followed by presentation of the thermochemical energy storage system concept, experimental results, discussion, conclusion and future work.

**BACKGROUND**

Wireless sensor modules have been proposed for deployment on the Moon and off-world environments. These sensors will be stationary, operate intermittently and hence consume minimal power. Several sensor modules can form a wireless sensor network. One such example of this concept is Space WIreless sensor networks for Planetary Exploration (SWIPE) [3]. The SWIPE project is identifying scientific experiments and communication technology that can be employed for exploration of the lunar surface. Typical experiments that come to mind include measuring variabilities in temperature, solar insolation, space weather and astronomy.

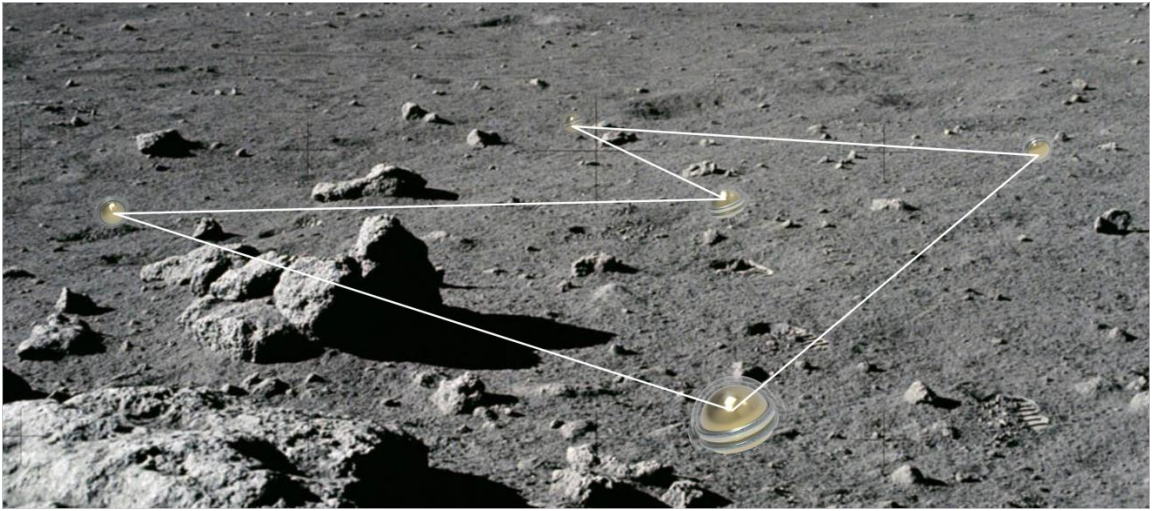

**Figure 1. Wireless Sensor Module Network.**

Power generation and thermal management remain the biggest challenges faced by planetary sensor network, as temperatures can vary significantly from day and night[4]. A sensor module was designed for off-world application at MIT's Field and Space Robotics Laboratory with the main goal of achieving a robust passive thermal design and exploring the feasibility of fuel cells as a long duration power supplies. This sensor module tested the feasibility of fuel cells for terrestrial and space applications. The design was based on that of a thermos wherein a vacuum is created between an inner sphere containing the electronics and outer sphere insulated by vacuum. However, a good vacuum was not achieved and hence the experiment did not give the expected temperature profile. The sensor module was tested at -24°C and reached steady state temperature within 4-6 hours.[4]

Thermochemical energy storage has been widely used in terrestrial applications for long term solar-energy storage. The system, open or closed contains a sorbent and a sorbate. Sorbates are liquids or gases like water vapor which upon reacting with sorbents releases heat. Using a combination of different heat exchangers, a cooling effect can be produced. During the summer, the solar irradiance charges the thermochemicals. Charging refers to external heating needed to evaporate sorbate from the sorbent. The discharging reaction takes place in winter wherein the stored solar energy is used to heat water. Discharging refers to release of heat energy when sorbent and sorbate come in contact.

**Material Selection Criteria[5]**

The selection of sorbents typically depends on number of factors. The crucial parameters are:

1. High water uptake
2. High specific energy
3. Low charging temperature.
4. Fast reaction kinetics
5. Availability
6. Ease of storage
7. Safety

Most pure chemicals (sorbents) have poor mass transfer characteristics and are poor conductors of heat. Pure chemicals can be prepared with composite materials to increase the effective thermal conductivity of the material and improve chemical reaction kinetics. They can overcome pressure drops across the salt bed and hence reduce energy required for regeneration purposes. Zeolites are class of compounds made of silicon and aluminum. The change in the ratio of silicon to aluminum alters the hydrophilic characteristics of the Zeolites.[6] This will increase the energy output as well as increase regeneration energy.

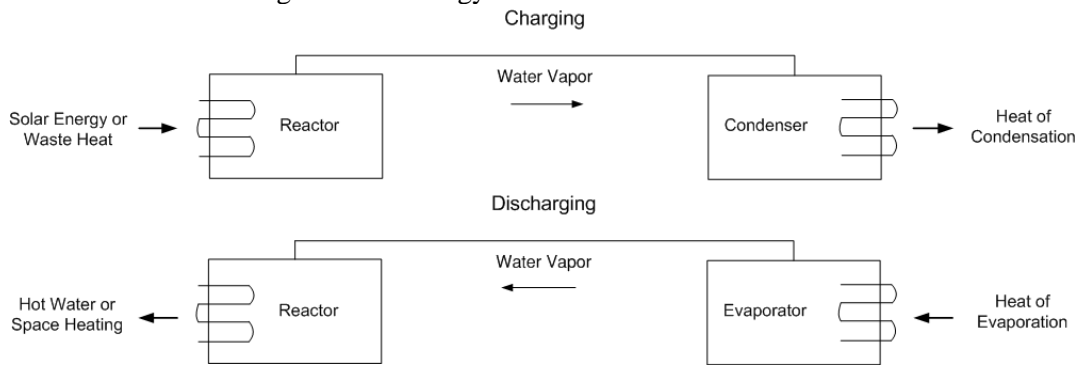

**Figure 2. Working of a Closed System.**

The user of thermochemicals for temperature control has attracted interest from the space community. A High-Capacity Spacesuit Evaporator Absorber Radiator (SEAR) for human space exploration has been proposed. The system was designed to provide a cooling effect, keeping the internal suit temperature at 20°C by rejecting heat with only radiation at 50°C[7]. The system con-sists of a lithium chloride heat rejection system coupled with a water membrane evaporator for heat gain. The system in essence is a portable vapor absorption system with heat input coming from solar energy.

**SYSTEM DESIGN**

The proposed system design for a sensor network module consists of two concentric spheres with the inner sphere housing the microprocessor, sensor electronics, radio and power supply[4]. This design was chosen as it minimizes heat loss. The inner sphere is divided into three parts. The top part contains all the electronics and the middle part contains the power source. The lower part is empty and is designed to keep reactants for the fuel cell. The inner sphere radius is 3.5 cm and the outer sphere radius is 5.5 cm.

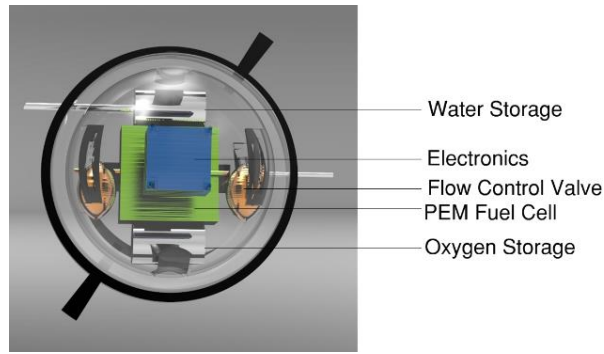

**Figure 4. Mission Concept Inner Sphere Design.**

This initial study is being conducted using batteries instead of fuel cells. The inner sphere is wrapped with a sheet of Aerogel insulation. This reduces the heat loss. The thermochemical is placed in a container surrounding the internal sphere.

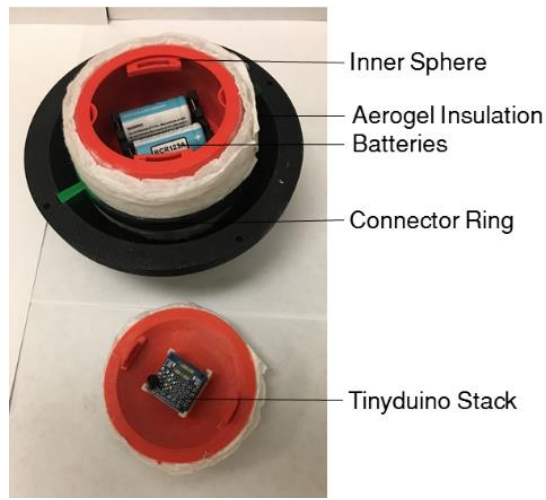

**Figure 5. System Layout.**

The microprocessor and all other electronic components, including the radio take up a 2 cm × 2 cm × 2 cm volume. Two temperature sensors, TMP 36 and BMA 250 are used to measure temperature.

**Conventional Thermal System and Thermochemical Energy Storage System.**

In a conventional thermal control system, the heating effect is provided by kapton heaters. Kapton heaters are used to heat critical subsystems rather than providing heating to the entire spacecraft. The block diagram for the conventional system is shown in Figure 6. A conventional thermal control system attempts to thermally isolate the internal systems from the external environment. This is most beneficial when the external temperatures plummet.

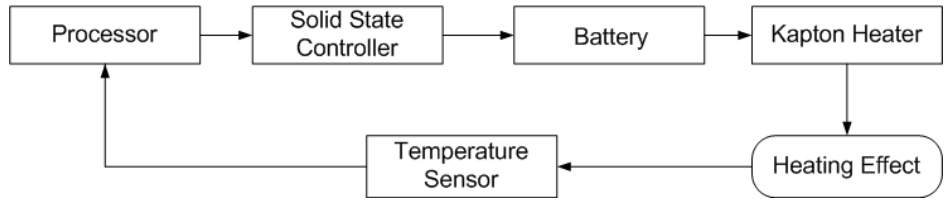

**Figure 6. Conventional Thermal Control System.**

TESS shown in Figure 7 captures external heat from the environment during daylight and releases it during the nighttime, when the external temperatures plummet. The feasibility of a passive thermal control system depends on controlling the reaction rate. This mechanism will hence control the rate of heat production. The system will have a flow control valve and a micro pump. Similar to the conventional system, a processor will check the temperature sensor data. In this system, temperature sensors will actuate the flow control valve. The water vapor flow onto the dehydrated salt bed and will give the necessary heating effect. Once the set point temperature is reached, the processor will stop the flow control valve.

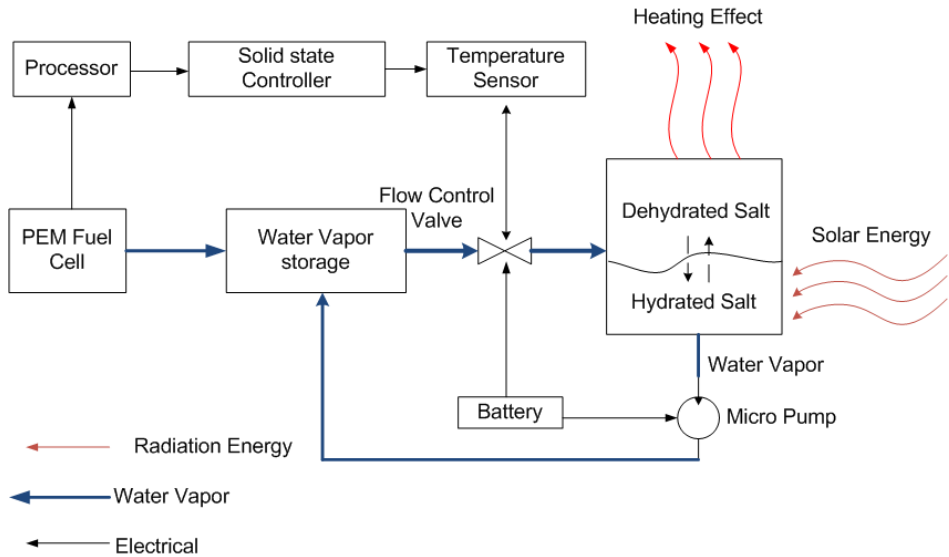

**Figure 7. Proposed Thermochemical Energy Storage System.**

**System Budgets**

The mass budget for our experimental system is shown below. The system power budget shown in Table 2 lists the power generated by each component within the system. All components reject heat to the environment due to operational inefficiencies. In the sensor module, all the electronics are low power devices. The lithium ion battery has a discharge efficiency of 85 % at room temperature. The low power electronics also have inefficiencies associated with them. A 10 % heat loss is assumed for all electronics. The SD card module consumes 0.35 W only when it records the data. Since data is being recorded at every 10 seconds, this value is assumed to be constant over the entire experiment.

**Table 1. Mass Budget.**

| Subsystem | Component | Mass(g) | Deviation | Max Mass (g) |
|---|---|---|---|---|
| Structure | Outer sphere | 178 | 1.2 | 210 |
|  | Inner Sphere | 102 | 1.2 | 120 |
|  | Connector ring | 5 | 1.1 | 6 |
|  | Battery Holders | 19 | 1.1 | 200 |
|  | TCM container | 21 | 1.1 | 23 |
| Command & data Handling | Microprocessor | 4 | 1.1 | 5 |
|  | SD card module | 4 | 1.1 | 5 |
| Communication | UHF Radio | 7 | 1.1 | 8 |
| Sensors | BMA 250 | 4 | 1.1 | 4 |
|  | TMP 36 | 2 | 1.1 | 2 |
| Power | Battery | 25 | 1.3 | 33 |
| Thermal | Insulation | 3 | 1.1 | 3 |
|  | TCM salt | 25 | 1.2 | 30 |
| Total |  |  |  | 470 |

**Table 2. Power Budget.**

| Component | Current(mA) | Voltage(V) | Power Consumed (mW) | Heat Lost (mW) |
|---|---|---|---|---|
| Processor | 1.2 | 3.5 | 4.2 | 0.42 |
| SD Card | 100 | 3.5 | 350 | 35 |
| Sensor board | 0.14 | 3.5 | 0.49 | 0 |
| TMP 36 | 0.05 | 3.5 | 0.175 | 0 |
| Battery |  |  | 355 | 53 |
| Total Heat Lost |  |  |  | 90 |

**Thermal Model**

Here we develop a thermal model of our experimental system and compare a spherical system versus a cube. The advantage of the steady-state model is that it is independent of mass and specific heat. The thermal design of the system without thermochemical storage system is shown. For a sphere, a 1D lumped capacitance model is created[9]. The model is based on a generic terrestrial version where convection exists. For our experiments, we ignore convection as it will have limited effect in our target environments. The solution at each node represents the temperature there and can be found in reference[10].

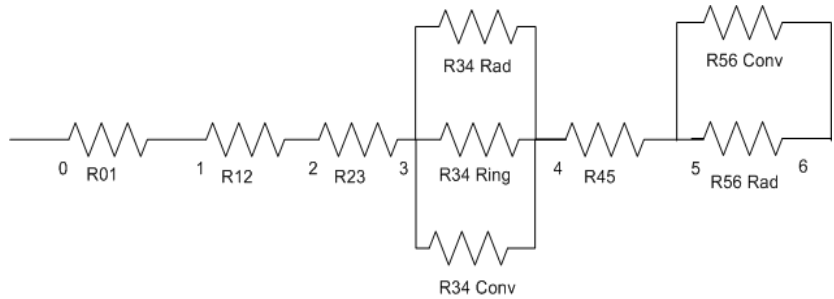

**Figure 8. Thermal Resistance Model.**

Based on the results of this analysis, a sphere in a sphere configuration was chosen over a cube in a cube configuration (Figure 10). A sphere has a higher volume to surface area ratio, thus reducing heat loss. The cube will have greater surface area.

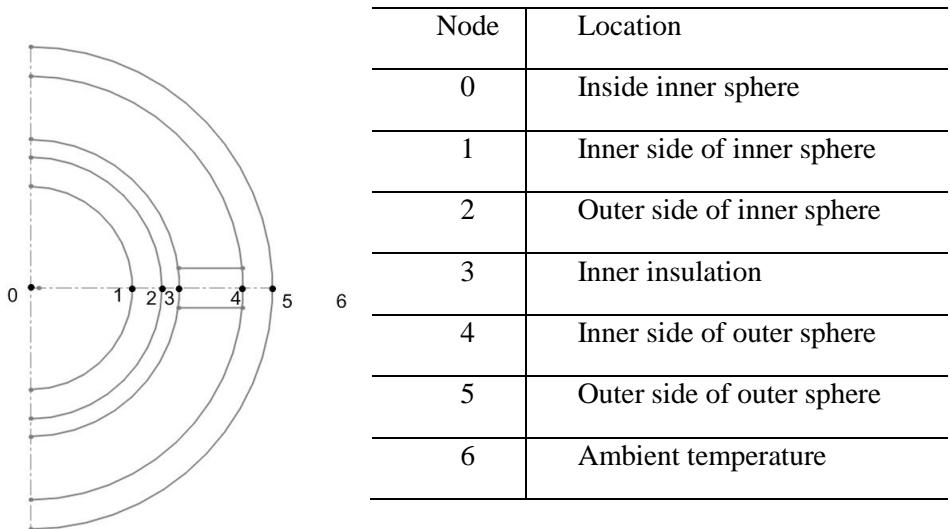

| Node | Location |
|------|----------|
| 0 | Inside inner sphere |
| 1 | Inner side of inner sphere |
| 2 | Outer side of inner sphere |
| 3 | Inner insulation |
| 4 | Inner side of outer sphere |
| 5 | Outer side of outer sphere |
| 6 | Ambient temperature |

**Figure 9. Position of Nodes.**

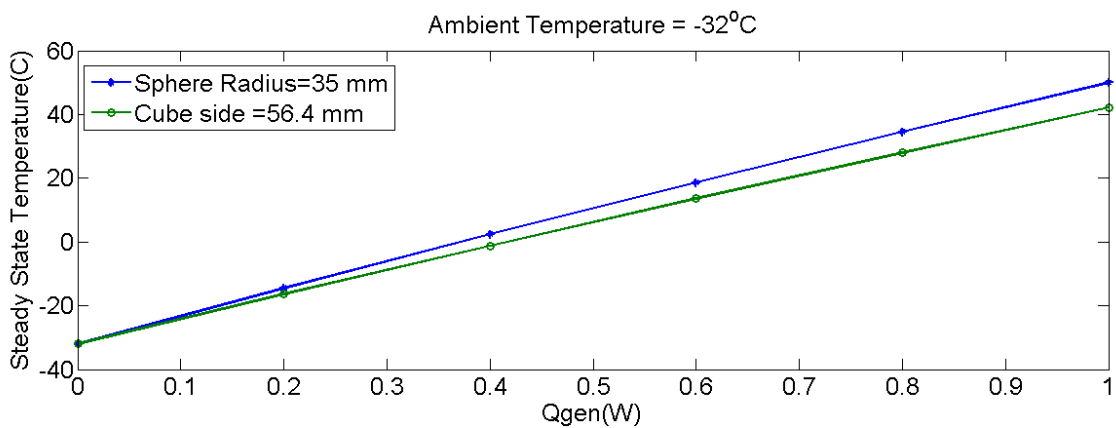
**Figure 10. Comparison of Sphere and Cube.**

**SORBATE SELECTION**

Having compared and identified the best geometry, the next task involves selecting right sorbate. The heat released by sorbate is a system property rather than a material property.[11] The maximum heat theoretically released can be known if the heat of reaction is known. The energy storage density at STP can calculated using, [12]

$$\Delta H_r = \Delta H_{fh} - (x\Delta H_W + \Delta H_{fd}) \qquad (1)$$

$$\Delta H_r = \sum H_{f_{products}} - \sum H_{f_{reactants}} \qquad (2)$$

**Table 3. Energy Storage Capacity.**

| Dehydrated Salt | $\Delta H_{fd}$ | Molar mass(g/mol) | Hydrated Salt | $\Delta H_{fh}$ | $\Delta H_r$ | Energy Storage (Wh/kg) |
|---|---|---|---|---|---|---|
| LiCl | -408 | 42.4 | LiCl.H$_2$0 | -712 | -56 | 370 |
| LiCl | -408 | 42.4 | LiCl.2H$_2$0 | -1013.7 | -109.7 | 720 |
| LiCl | -408 | 42.4 | LiCl.3H$_2$0 | -1311 | -159 | 1050 |
| LiCl | -408 | 42.4 | LiCl.5H$_2$0 | -1889.11 | -241.11 | 1600 |
| MgSO$_4$ | -1278 | 120.36 | MgS0$_4$.7H$_2$0 | -3388 | -374 | 870.0 |
| MgCl$_2$ | -641 | 95.21 | MgCl$_2$.6H$_2$0 | -2499 | -370 | 1100 |
| SrBr$_2$ | -717 | 247.4 | SrBr$_2$.6H$_2$0 | -2531 | -326 | 370 |
| CaCl$_2$ | -795 | 110.98 | CaCl$_2$.6H$_2$0 | -2607 | -363.2 | 900 |

Some of the commercially available sorbents were considered. As lithium chloride has a high theoretical energy density and also has space heritage, it was chosen for our experiments. Lithium chloride forms multiple hydrates. The penta hydrate have the highest energy storage and is stable at very low temperatures of approximately -80 ºC [12]. Experiments were conducted at -32ºC.

**EXPERIMENTS**

Figure 11 shows the experimental setup. The experimental results are plotted below in Figure 12 and 13. The heater is programmed to maintain the temperature at -20°C. The graph shows that at -20°C, the heater initially maintains the temperature at -20 °C before it drops down to the ambient.

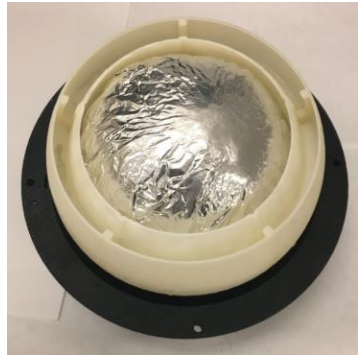

Figure 11. Experimental Setup for Sensor Module.

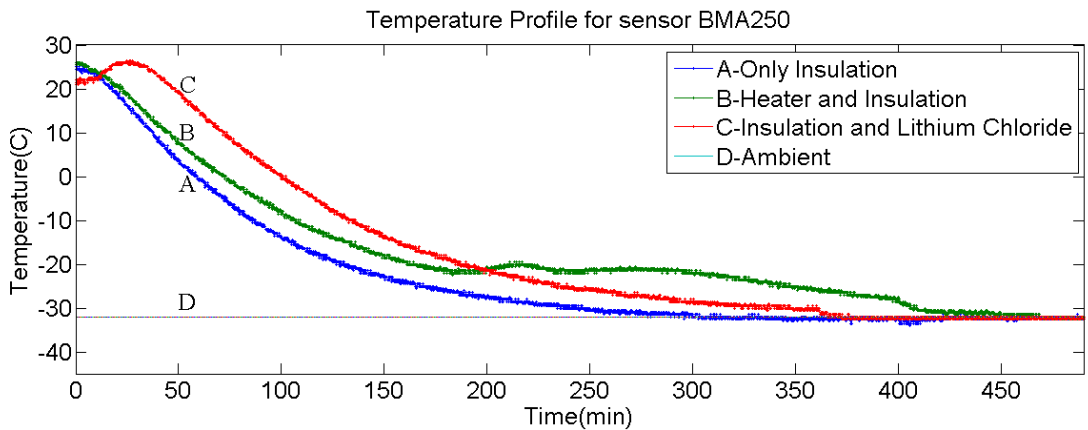

Figure 12 .Comparison of experimental temperature profiles at -32°C for sensor BMA250.

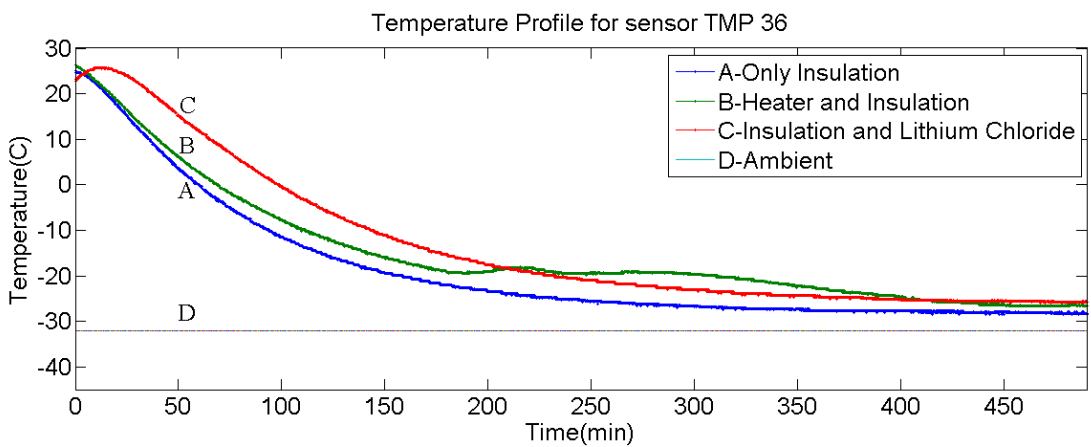

Figure 13 .Comparison of experimental temperature profiles at -32°C for sensor TMP 36.

Table 4. Experimental Results.

| Ambient Temperature=241K (-32°C) | | | | | | |
|---|---|---|---|---|---|---|
| | A –TMP 36 | A-BMA | B -TMP 36 | B -BMA | C- TMP 36 | C-BMA |
| Steady State Temp(C) | -28 | -32 | -26 | -32 | -26 | -32 |
| Time(min) | 383 | 315 | 448 | 450 | 412 | 419 |

**Repeatability Experiments**

The experiment was repeated four times with similar results recorded (Figure 14, 15). The BMA 250 records -33°C in one of the experiments. This is can be attributed to the inaccuracy of the sensor and deviation in the refrigerator temperature.

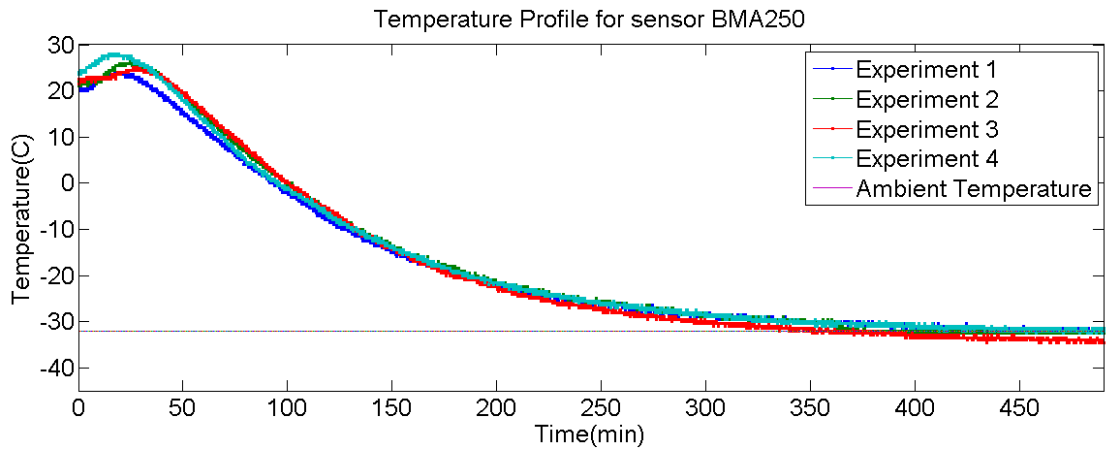

Figure 14 . Repeatability experiments at -32°C. Profiles for BMA 250 sensor.

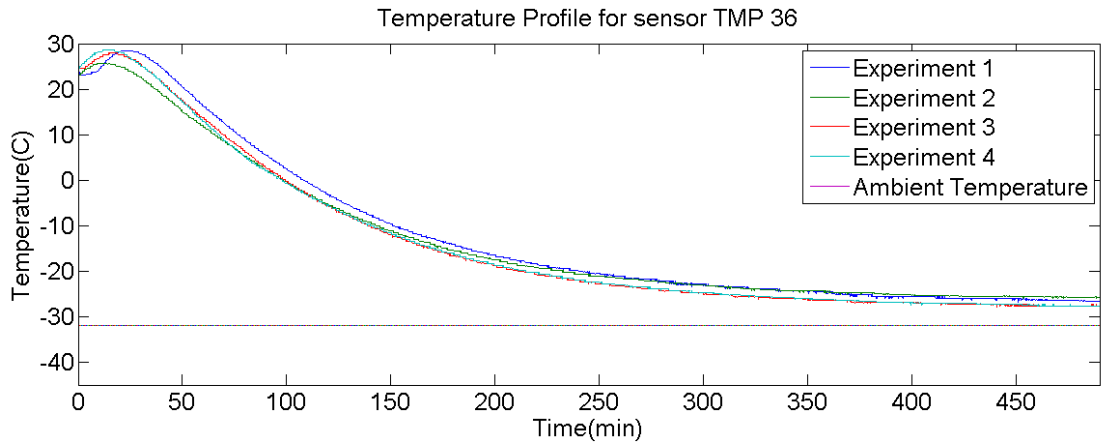

Figure 15. Repeatability experiments at -32°C . Profiles for TMP 36 sensor.

Table 5. Repeatability Experiment Results.

| Experiment run | 1 | | 2 | | 3 | | 4 | |
|---|---|---|---|---|---|---|---|---|
| Ambient Temperature=241K (-32$^0$C) | | | | | | | | |
| | LiCl TMP | LiCl BMA | LiCl TMP | LiCl BMA | LiCl TMP | LiCl BMA | LiCl TMP | LiCl BMA |
| Steady State Temp(C) | -26 | -32 | -26 | -32 | -28 | -33 | -27 | -32 |
| Time(min) | 412 | 419 | 422 | 411 | 414 | 411 | 408 | 411 |

The results show that the proposed thermochemical energy storage system is an improvement over insulation alone and is even better than electrical heating. The problem with electrical heating is that it consumes valuable energy that would otherwise be used to power the sensor electronics. However, results show that the sensor cannot keep the sensor networks above storage temperatures of -20 $^o$C beyond 3 hours. The system as is maybe applicable on asteroid and small bodies with short rotational periods of 3 hours or less.

## DISCUSSION

A thermochemical storage system has been proposed to solve the heating challenges faced by sensor modules in cold temperatures. Using 25 grams of lithium chloride and 25 grams of water, the system may be held at storage temperatures for about 3 hours. For the system to survive for longer duration requires more thermochemicals. The thermochemical approach show a clear advantage over insulation alone and even electrical heating which consumes valuable energy. The experiments showed good repeatability with deviation in results of 4%. This deviation is due primarily to inaccuracy of the sensor, inaccuracy of the refrigerator and human error.

There is no direct comparison between the performance of the heater and the TESS. Area under the curve represents only an estimate of the change in energy of the system. The area under the curve for the heater was less than the TESS for both experiments. For TESS the value is 2777 C.min and for heater, the value is 1901C.min. It shows that LiCl released more energy than the kapton heaters.

It should be noted that this increase was achieved using freezing water and not vapor. In case of water, there is a greater chance of a crystalline layer being formed on the surface of the salt bed. It is plausible that this phenomenon reduced further diffusion of water, thus reducing the magnitude of the heat released. With water vapor, there should be better diffusion of water molecules through the bed. With proper design it should be possible to improve the performance of the system further.

The main advantage of the proposed system is the potential of developing a rechargeable system. This system would ideally need only the waste water from a fuel cell as input to provide heat. Although this hasn't been demonstrated yet, these preliminary experiments offer a promising pathway towards our next set of experiments.

## CONCLUSIONS

We have developed a thermal control system that utilizes chemicals to store and dispense heat when needed. The system takes waste products, such as water from a sensor module and transfers them to a thermochemical storage system. These thermochemical storage systems when mixed with water (a waste product from a PEM fuel cell power supply) releases heat. Under eclipse, the heat from the thermochemical storage system is released to keep the probe warm enough to survive the night. The results point towards a promising pathway towards application of this technology towards the Moon, Mars and asteroids. Next steps will involve developing an improved sensor module that survive Mars equatorial surface conditions.